\documentclass[twocolumn]{aastex61}
\begin{document}

\title{Weighing Uranus' moon Cressida with the $\eta$ ring}

\author[0000-0002-7867-7674]{Robert O. Chancia}
\affiliation{Department of Physics, University of Idaho, Moscow ID 83844-0903}

\author{Matthew M. Hedman}
\affiliation{Department of Physics, University of Idaho, Moscow ID 83844-0903}

\author{Richard G. French}
\affiliation{Astronomy Department, Wellesley College, Wellesley MA 02481}

\correspondingauthor{Robert O. Chancia}
\email{rchancia@uidaho.edu}

\begin{abstract}
The $\eta$ ring is one of the narrow rings of Uranus, consisting of a dense core that is 1-2 km wide and a diffuse outer sheet spanning about 40 km. Its dense core lies just exterior to the 3:2 Inner Lindblad Resonance of the small moon Cressida. 
We fit the $\eta$ ring radius residuals and longitudes from a complete set of both ground-based and Voyager stellar and radio occultations of the Uranian rings spanning 1977-2002. We find variations in the radial position of the $\eta$ ring that are likely generated by this resonance, and take the form of a 3-lobed structure rotating at an angular rate equal to the mean motion of the moon Cressida. The amplitude of these radial oscillations is $0.667\pm0.113$ km, which is consistent with the expected shape due to the perturbations from Cressida. The magnitude of these variations provides the first measurement of the mass and density of the moon Cressida ($m=2.5\pm0.4\times10^{17}$ kg and $\rho=0.86\pm0.16$ g/cm$^3$) or, indeed, any of Uranus' small inner moons. A better grasp of inner Uranian satellite masses will provide another clue to the composition, dynamical stability, and history of Uranus' tightly packed system of small moons.
\end{abstract}


\section{Introduction}
In March of 1977, \citet{1977Natur.267..328E}, \citet{1977Natur.267..330M}, and \citet{1977Natur.267..331B} discovered nine narrow rings around the planet Uranus by measuring the light blocked by each ring before and after Uranus occulted the star SAO158687. Since then, the Uranian rings have been studied extensively with ground based stellar occultations \citep{1978AJ.....83..993M,1978AJ.....83.1240N,1981AJ.....86..127E,1981AJ.....86..444E,1981AJ.....86..596N,1982Natur.298..827F,1982Icar...52..454S,1983Icar...56..202E,1984AJ.....89.1587E,1986Sci...231..480F,1986Icar...67..134F,1987Icar...71...91E,1988Icar...73..349F,1996Icar..119..269F}. Occultations provide very precise radial locations of the rings at different longitudes in their orbits around Uranus. \citet{1988Icar...73..349F} found that the main rings of Uranus consist of six measurably eccentric rings ($6$, $5$, $4$, $\alpha$, $\beta$, and $\epsilon$) and three nearly circular rings ($\eta$, $\gamma$, and $\delta$). 
In the past, measurements of the $\eta$ ring's radius have not shown the ring to be anything but circular. The $\eta$ ring also features a broad low optical depth sheet extending approximately $40$ km exterior to its narrow core \citep{1983Icar...56..202E}. 

During the Voyager 2 flyby of Uranus \citet{1986Sci...233...43S} discovered ten new small inner moons, but no one has ever measured their masses or densities. Nine of the moons orbit within a radial range of $20$,$000$ km, making the group one of the most tightly packed systems of interacting satellites in our solar system. \citet{1995Icar..114..217L} estimated the masses of the inner moons assuming densities equal to that of the larger moon Miranda \citep{1992AJ....103.2068J} and shapes estimated with photometry \citep{1989Icar...81...92T}, but stated that at least some of Uranus' small inner moons are significantly less massive than these estimates. The lifetime of this system is highly sensitive to the masses of the individual satellites \citep{2015AJ....149..142F}. In fact, prior to the knowledge of the even less stable moon Cupid \citep{2003IAUC.8209....1S,2012Icar..220..911F}, \citet{1997Icar..125....1D} showed that Desdemona could collide with either Cressida or Juliet within the next $4-100$ million years, depending on the masses of the satellites involved. The discovery of the dusty $\nu$ and $\mu$ rings \citep{2006Sci...311..973S}, near the orbits of Portia/Rosalind and Mab respectively, hints at the possibility of an evolving inner ring-moon system dominated by accretion \citep{2013ApJ...765L..28T}. \citet{2015Icar..254..102K} also argue that anomalies in Mab's orbital motion may be explained by a ring-moon system that is undergoing re-accretion after a recent catastrophic disruption.

Here we investigate a complete set of Uranian $\eta$ ring occultation observations spanning their discovery in 1977 to 2002. 
We find that the $\eta$ ring's radii exhibit a 3-lobed structure rotating around Uranus at the mean motion of the moon Cressida. We argue that this structure is a result of the $\eta$ ring's close proximity to Cressida's 3:2 inner Lindblad resonance (ILR). One of the maxima in the ring's radius aligns with Cressida, as expected for the stable ring structure located exterior to the resonant radius. The measured radial amplitude of this ring structure and its distance from the resonance allow us to estimate Cressida's mass, and thus obtain the first gravity-based mass measurement of any inner Uranian moon.

We have only been able to find three previous mentions of the Cressida 3:2 ILR and its association with the $\eta$ ring. \citet{1987AJ.....93..724P} identified the most relevant resonances in the Uranian ring-moon system and made a case for Cordelia and Ophelia shepherding the outermost $\epsilon$ ring through torques generated by the Lindblad resonances located appropriately on the ring's inner and outer edges \citep{1987AJ.....93..730G}. They also note single resonances that could be perturbing the $\gamma$ and $\delta$ rings. Finally, they state: ``The only isolated first-order satellite resonances which fall near any of the remaining rings are located interior to the $\eta$ ring." \citet{1987AJ.....93..724P} list both the Cressida 3:2 and the Cordelia 13:12 resonances, located at $a=47171.6\pm0.3$ km and $a=47173.0\pm0.3$ km respectively. These resonances fall $3-5$ km interior to the $\eta$ ring. They calculate the widths of both resonances to be $\sim 1$ km and dismiss the possibility that either resonances is perturbing the $\eta$ ring. \citet{1988VA.....32..225M} later marked the location of the Cressida 3:2 ILR in their figure displaying a radial scan of a high phase image of the Uranian rings acquired by Voyager 2. Subsequently, \citet{1990Natur.348..499M} noted that this resonance needs to be re-examined using updated satellite parameters. At the time, with a smaller data set, there was no detection of either an $m=3$ or an $m=13$ mode in the $\eta$ ring, nor any other modes due to resonances with known satellites having observed effects on any of the other previously noted rings' edges \citep{1988Icar...73..349F}. Thus, it was only sensible to dismiss these resonances, and it is reasonable that they have not been of interest since. We are only able to make this discovery now because we have a larger set of occultation data extending from 1977 through 2002.

We present the data used in this analysis in Section 2, and describe our ring particle streamline model and our mode detection methods in Section 3. In Section 4, we report the parameters of our fit to the $\eta$ ring and calculate the mass and density of Cressida. Finally in Section 5, we discuss potential implications for the dynamical stability of the tightly packed system of inner Uranian moons and the possible composition of Cressida.

\section{Observational Data}
The observational data used for this analysis consist of 49 individual occultation observations of the $\eta$ ring. In the appendix, Table \ref{occdata} contains each occultation's ring intercept time, inertial longitude, and mid-radius determined using a simple square-well model for profile fitting, developed by \citet{1984AJ.....89.1587E} and used in later orbit determinations of the Uranian rings \citep{1986Sci...231..480F,1986Icar...67..134F,1988Icar...73..349F,1991uran.book..327F}. Of these 49 observations, 46 are Earth-based stellar occultations, two are Voyager 2 Radio Science Subsystem (RSS) radio occultations, and one is a Voyager 2 Photopolarimeter Subsystem (PPS) stellar occultation. Several of the observations are ingress and egress pairs from the same occultation of Uranus and its rings.

For each Earth-based occultation, an instrument recorded the brightness of the background star as a function of time. As the Earth moves relative to Uranus the rings can block the star's light, leaving each ring's mark as a sharp decrease in the recorded brightness of the star for some amount of time related to the width of the ring. Typically the observations were detected with an InSb photometer in the $2.2$ $\mu$m band, using the K filter, where Uranus is fainter than the rings. Most observations provided limited information about the radial structure within the rings, and here we are making use only of the estimate of the radius of the mid-point of each ring occultation profile. Interested readers should see \citet{1979ARA&amp;A..17..445E} for a review of stellar occultation studies of the solar system and \citet{1984prin.conf...25E} for a review of this observation method specific to the rings of Uranus.

To identify possible Uranus occultation opportunities \citet{1973JBAA...83..352T} compared positions of Uranus to stellar positions in the Smithsonian Astrophysical Observatory (SAO) catalog. Once the rings were discovered, it became more appropriate to utilize dimmer stars that are bright in the $2.2$ $\mu$m band. Thus, \citet{1977AJ.....82..849K} searched for stars on photographic plates containing star fields ahead of Uranus and created a list of ideal future occultation observations. Additional lists of this type were compiled by \citet{1981AJ.....86..138K}, \citet{1982AJ.....87.1881M,1985AJ.....90.1894M}, \citet{1988AJ.....95..562N}, and \citet{1991AJ....102..389K}. 

The Voyager 2 PPS stellar occultation only detected the $\eta$ ring on egress \citep{1986Sci...233...65L,1990Icar...83..102C}. In the case of the Voyager 2 RSS occultations, the RSS instrument illuminated the rings at $3.6$ cm and $13$ cm wavelengths in the direction of Earth once beyond the ring plane. Stations on Earth detected the diffracted signal and relative phase change, to later be reconstructed into high-resolution radial optical depth profiles after the removal of diffraction effects  \citep{1986Sci...233...79T,1989Icar...78..131G}. Presently, ground based occultation opportunities are rare because Uranus has passed out of the dense Milky Way background, drastically reducing the density of appropriate background stars. The rings are also no longer as open to our view from Earth as they were in the 1980s because the apparent aspect of the ring plane as viewed from Earth changes over time.


\section{Ring Particle Streamline Model and Fitting Method}
The procedure used here follows that of \citet{1986Sci...231..480F,1988Icar...73..349F,1991uran.book..327F} for the Uranian rings, more recently employed by \citet{2010AJ....139..228H}, \citet{2014Icar..227..152N,2014Icar..241..373N}, and \citet{2016Icar..274..131F} for analyses of Saturn's non-circular narrow rings, gaps, and edges. After taking account any inclination relative to the equatorial plane, the majority of narrow rings are well-fit by simple precessing Keplarian ellipses whose radii are described by:
\begin{equation} \label{keprad}
r(\lambda,t)=\frac{a(1-e^2)}{1+e\cos f},
\end{equation}
where the true anomaly $f=\lambda-\varpi_0-\dot{\varpi}(t-t_0)$. Here, the radius of the ring will vary with longitude $\lambda$ and time $t$, where $a$ and $e$ are the ring's semi-major axis and eccentricty, $\varpi_0$ is the ring's longitude of periapsis at the time $t_0$, and $\dot{\varpi}$ is the ring's apsidal precession rate. We can approximate a nearly circular ($e\simeq 0$) ring's radii as $r\simeq a(1-e\cos f)$. 

\begin{figure}[t]
\plotone{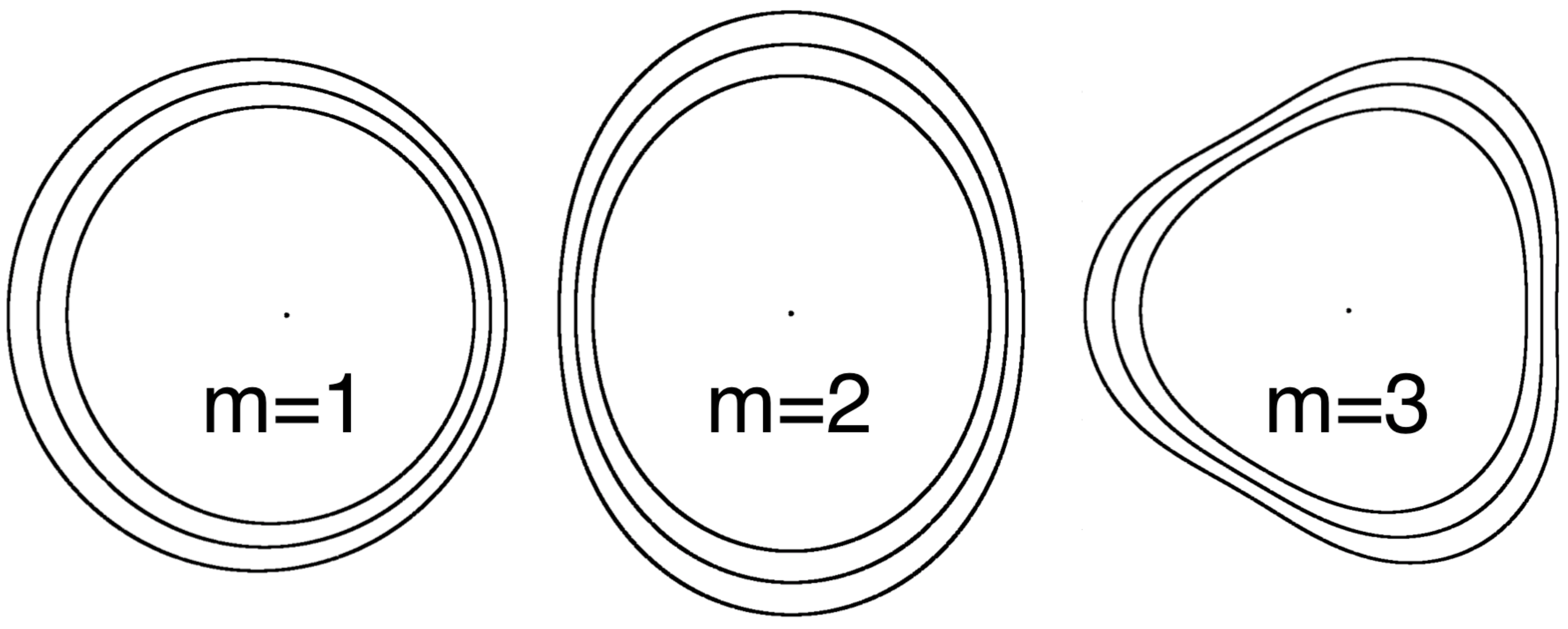}
\caption{The above shapes are an exaggerated representation of the $m$-lobed ring streamlines we detect in the Uranian rings. For each case of $m$, we have shown 3 streamlines with slightly different semi-major axes and a positive eccentricity gradient. Our addition of an eccentricity gradient results in a narrower ring width at periapsis, as is the case for several of the Uranian rings. \label{streamlines}}
\end{figure}

Additionally, several rings are found to contain forced radial oscillations and in a few cases there are even rings whose structure is dominated by free normal mode oscillations. In these cases, the structures are distinct from circles or ellipses and their radii are described by:
\begin{equation} \label{modrad}
r(\lambda,t)\simeq a-A_{m}\cos(m\theta),
\end{equation}
where $\theta=\lambda-\Omega_{p}(t-t_{0})-\delta_{m}$, following the formalism of \citet{2014Icar..227..152N,2014Icar..241..373N} and \citet{2016Icar..279...62F}. Here, the systematic radial oscillations of the rings form a $m-$lobed figure rotating around their planet at a pattern speed $\Omega_{p}$ with a radial amplitude $A_m$ and phase $\delta_m$. We show some exaggerated models of $m$-lobed ring streamlines, resulting from both free normal modes and Lindblad resonances, in Figure \ref{streamlines}. While individual particles follow normal elliptical orbits, described by Equation \ref{keprad}, the ring as a whole consists of streamlines with $m$ azimuthally symmetric radial minima and maxima rotating around the planet with the frequency
\begin{equation} \label{patspeed}
\Omega_p\simeq \frac{(m-1)n+\dot{\varpi}_{sec}}{m}.
\end{equation}
Here, the mean motion $n$ and apsidal precession rate $\dot{\varpi}_{sec}$ are functions of the semi-major axis $a$ of the ring, and $m$ can be any positive or negative integer. If we consider the case of $m=1$ we find that $\Omega_p=\dot{\varpi}_{sec}$, $A_1=ae$, and $\delta_1=\varpi_0$, so that $r$ is equivalent to the approximation of Equation \ref{keprad} above. 

In the case of a free normal mode oscillation, the pattern speed will be equal to the expected pattern speed obtained by evaluating Equation \ref{patspeed} at the semi-major axis of the ring. However, if the ring is perturbed by a satellite through a first-order Lindblad resonance, then the ring structure will have a forced pattern speed matching the mean motion of the perturbing satellite $n_s=\Omega_p$ and will differ from the expected pattern speed slightly based on the ring's separation from the exact radius of the resonance $|a-a_{res}|$. The ring is perturbed by the satellite due to the near commensurate ratio of the ring particles' orbital periods and the period of the perturbing satellite. As such, first-order Lindblad resonances are defined by $|m|:|m-1|$, where for every $|m|$ orbits of the ring particle, there are $|m-1|$ orbits of the corresponding satellite. In the majority of cases, the perturbing satellite lies at a larger semi-major axis than the ring ($a_s>a$). The relevant resonances in this case are called inner Lindblad resonances (ILR) and are assigned positive values of $m$. In the rare case of a satellite located interior to the rings it is possible to have both ILR and outer Lindblad resonances (OLR) at locations within the rings, allowing for negative values of $m$.

\begin{figure}[t]
\plotone{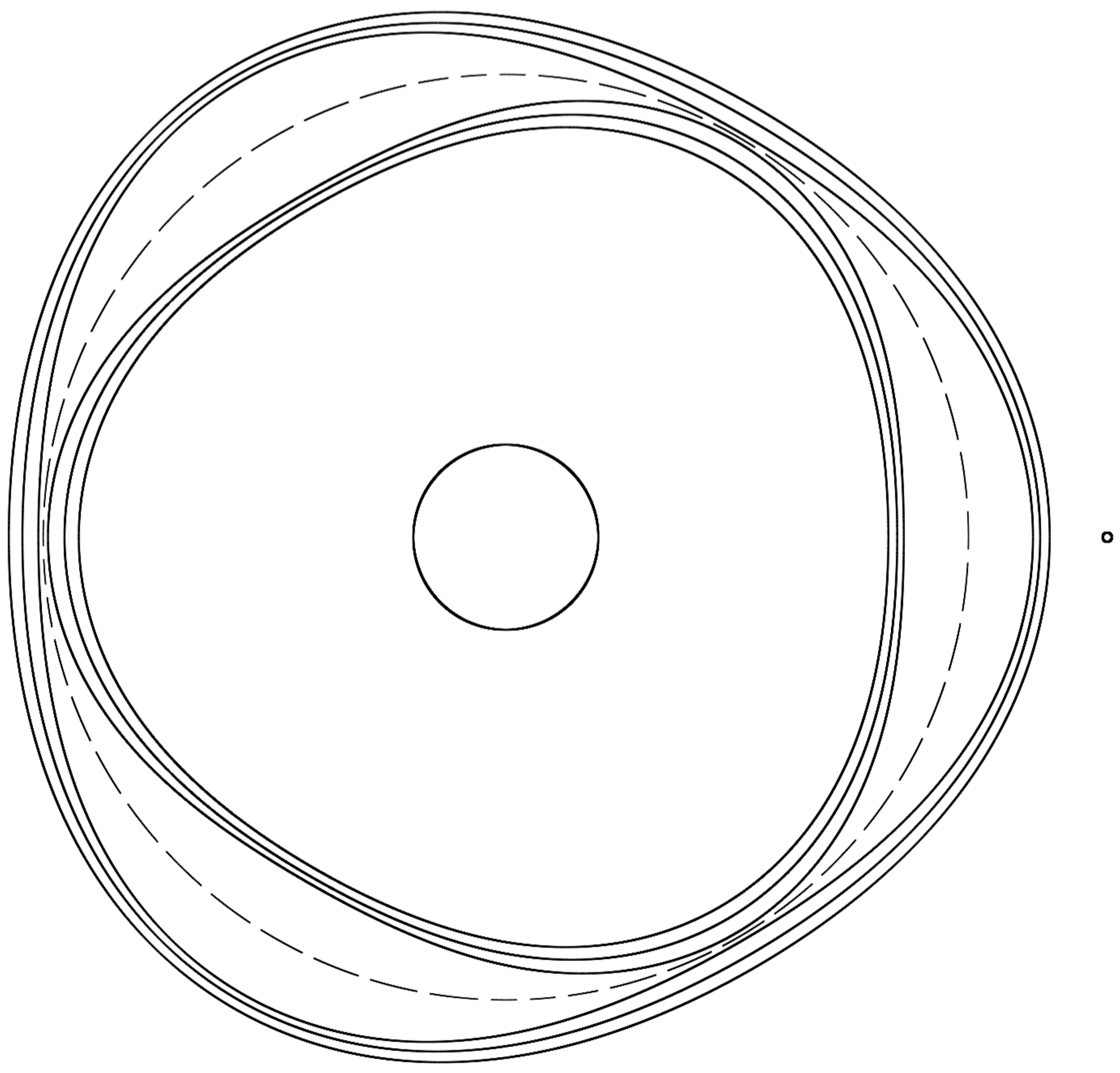}
\caption{An exaggerated cartoon model of ring particle streamlines around a planet and near a 3:2 ILR with an exterior moon, in the co-rotating frame of the moon. We've marked the resonant radius with a dashed line and included three ring particle streamlines on either side of the resonance. This shows the stable configuration on either side of the resonance, where a periapse (apoapse) is aligned with the moon interior (exterior) to the resonance. \label{3:2model}}
\end{figure}

The condition for a first-order Lindblad resonance is that the resonant argument:
\begin{equation} \label{resarg}
\varphi=m(\lambda-\lambda_s)-(\lambda-\varpi)
\end{equation}
is constant in time. Here $\lambda$ and $\lambda_s$ refer to the longitudes of a ring particle and the satellite respectively and $\varpi$ is the longitude of periapsis of the ring particle. If we consider a conjunction of the ring particle and the satellite ($\lambda-\lambda_s=0$) occurring when the ring particle is also located at its longitude of periapsis ($\lambda-\varpi=0$), then the condition that $\varphi$ is constant implies that all future conjunctions will occur when the ring particle is near periapsis. In general, this means that the ring particle will always be in the same phase of its orbit when it passes longitudinally close to the satellite. This allows the perturbing satellite to force the eccentricity and periapsis locations of streamlines located near the resonance. In Figure \ref{3:2model} we show a cartoon model of the resulting streamlines surrounding a 3:2 ILR in the co-rotating frame of the perturbing satellite. Interior (exterior) to the resonant radius, marked with the dashed line, the streamlines are stable when oriented such that one of the three periapses (apoapses) is aligned with the satellite.

In short, our procedure is a search for patterns in the varying mid-radii measurements of the rings. Each ring occultation observation provides the ring's radius at a particular longitude and time. To search for patterns in each ring we need the observed parameters, an $m$ value to test, and the resulting expected pattern speeds for that $m$ value. For each test of $m$, we compute the expected patten speed for the semi-major axis of the ring using Equation \ref{patspeed} and create an array of $100,000$ pattern speeds, evenly spaced in increments of $0.00001^{\circ}/$day, surrounding the expected pattern speed. Using each pattern speed we calculate $m\theta$, for every ring observations' longitude $\lambda$ and time $t$, using an initial epoch time $t_0$ of UTC 1977 MAR 10 20:00:00.00. We can then compute the observed ring radii $r$ vs. $m\theta$ mod $360^{\circ}$ and fit the data to a single sinusoid. The resulting fit parameters are $a$, $A_m$, and $\delta_m$, allowing us to compute model values of $r$ using Equation \ref{modrad}. We compute the RMS deviation of the observed radii and the model radii for each $m$'s $100,000$ test pattern speeds and look for a RMS minimum to identify the best fitting pattern speed. 

\begin{figure*}[t]
\plottwo{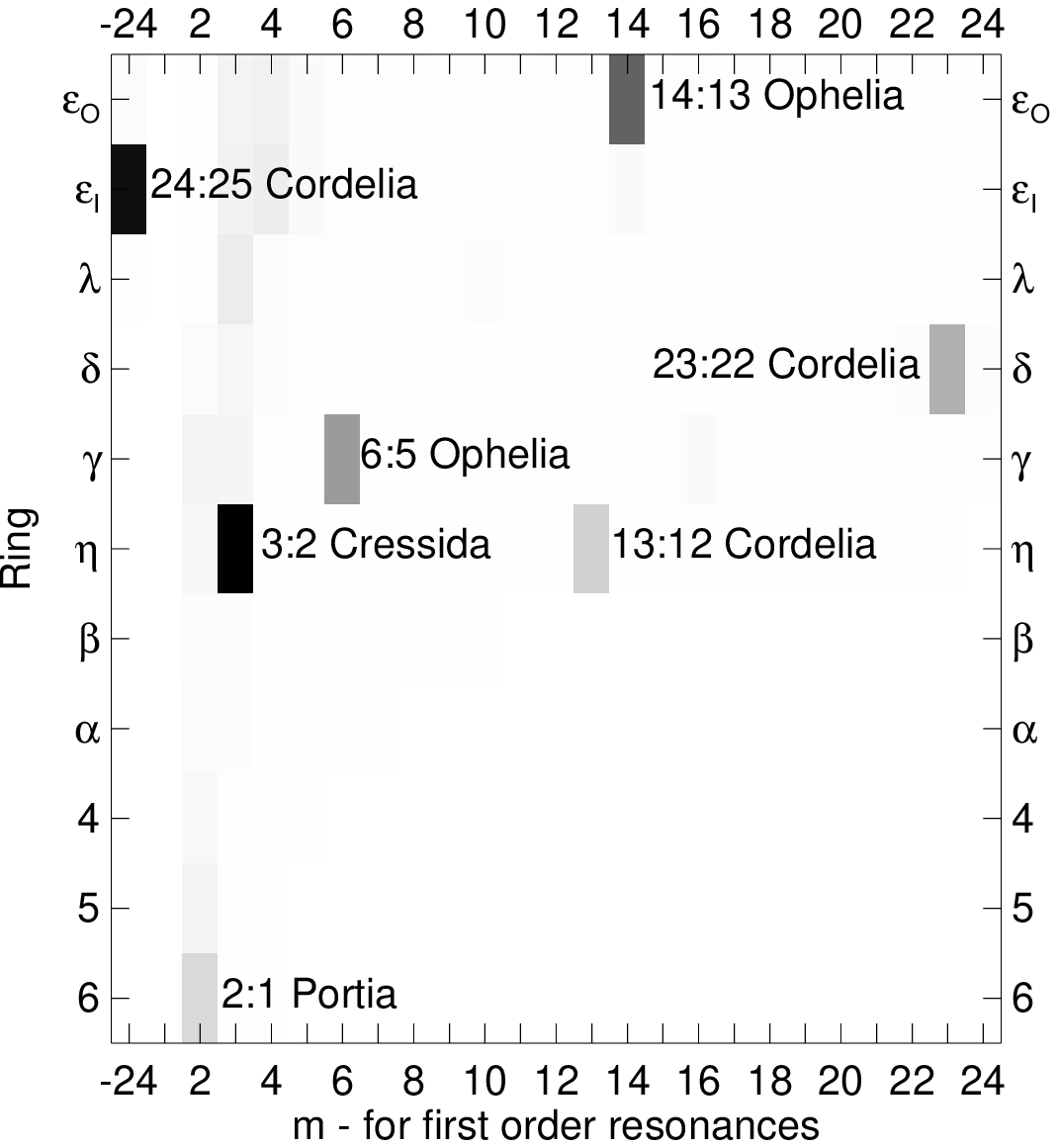}{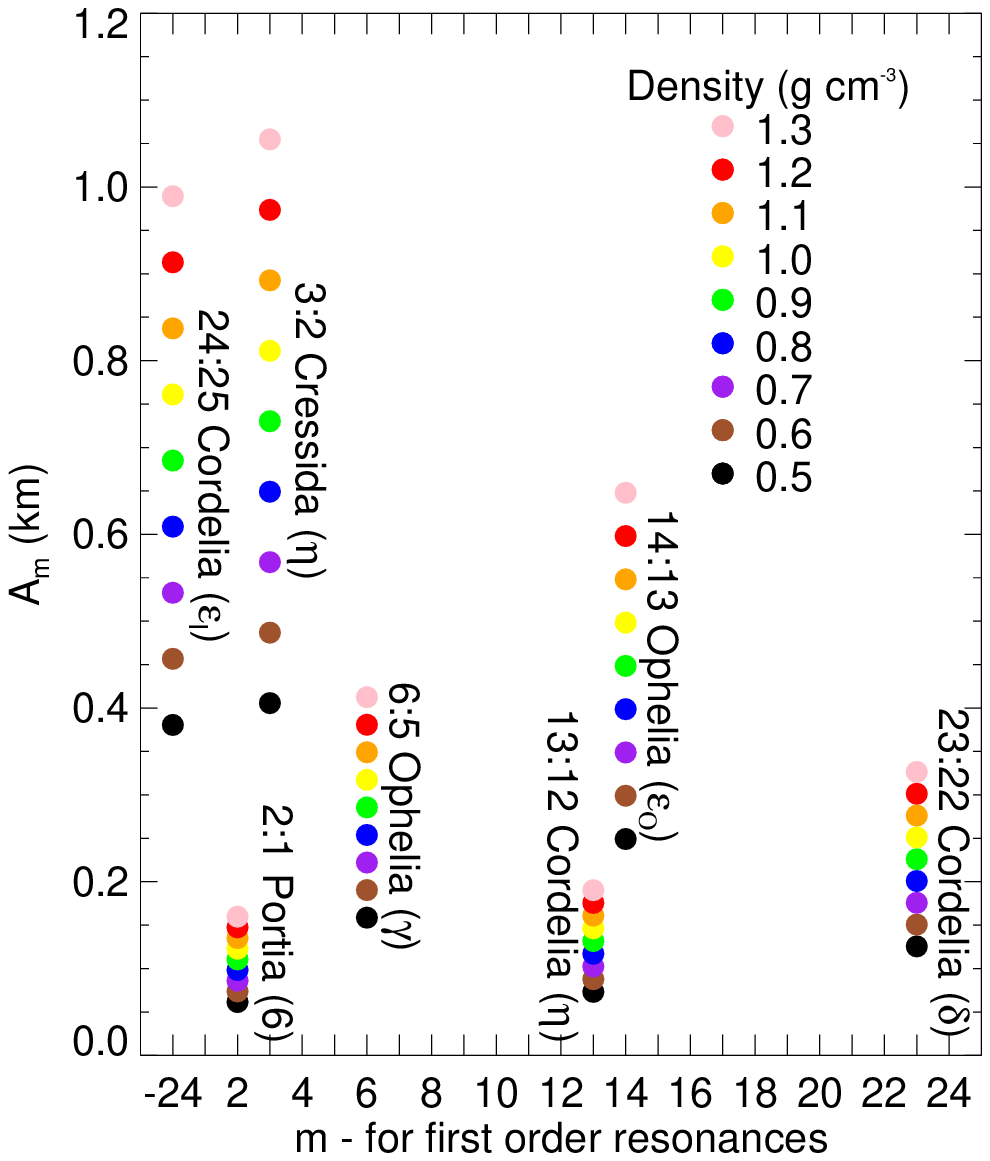}
\caption{The left mosaic shows the relative forced amplitude (darker = larger amplitude) of all first-order resonances of inner moons out to Perdita interacting with the Uranian rings assuming the moons each have a density of $1.3$ g/cm$^3$. The largest amplitude resonances are labeled, while the fainter patches in the mosaic are evidence of resonances within the system that do not fall close enough to any rings and would have much smaller amplitudes. The actual values of the significant resonances are plotted on the right side to compare with the mosaic, but with the addition of a range of moon densities ($0.5$ to $1.3$ g/cm$^3$), calculated using Equation \ref{MD10.22}. \label{resstrength}}
\end{figure*}

We first checked our algorithms by searching for known structures in the Uranian rings. In several rings ($6, 5, 4, \alpha, \beta,$ and $\epsilon$) we can easily detect RMS deviations that drop to nearly zero (sub-km) with the proper pattern speed and $m$ input. These are the rings that largely follow classical Keplerian ellipses ($m=1$) and whose pattern speeds equal the rings' apsidal precession rate, $\Omega_{p}=\dot{\varpi}_{sec}$. The $\eta, \gamma,$ and $\delta$ rings are nearly circular and their residuals are relatively larger when fit with a low amplitude $m=1$ ellipse. 
We are also able to identify the known $m=2$ structure of the $\delta$ ring and the combination of $m=0$ and $m=1$ for the $\gamma$ ring \citep{1986Sci...231..480F}.

We decided to identify the strongest resonances in the Uranian rings to have a better idea of the resonantly forced modes that are the most likely to be detected. To quantify the `strength' of the resonances in the system we chose to compare the expected forced radial amplitude on rings near each of the possible resonances in the main ring system. We use Equation 10.22 from Chapter 10 of \citet{1999ssd..book.....M},
\begin{equation} \label{MD10.22}
A_m=\frac{2\alpha a^2(m_{s}/m_{p})|f_d|}{3(j-1)|a-a_{res}|}
\end{equation}
where $A_m$ is the forced radial amplitude of a ring particle in the vicinity of a Lindblad resonance \citep{1987Icar...72..437P,1982ARA&amp;A..20..249G}. This amplitude is a function of the ratio of the perturbing satellite and central planet masses $m_s/m_p$, the radial separation of the ring and the resonance $|a-a_{res}|$, the ratio of the ring and satellite semi-major axes $\alpha=a/a_s$, and the Laplace factor $f_d$, that depends on $j$, the integer coefficient of the satellites longitude in the resonant argument, which is equivalent to $m$ in the case of a first-order Lindblad resonance. As shown in Figure 10.10 of \citet{1999ssd..book.....M}, $\frac{2\alpha|f_d|}{j-1}$ varies between $1.5$ and $1.6$, depending on $j$. Note that Equation \ref{MD10.22} isn't necessarily applicable for all cases. If $|a-a_{res}|$ is smaller than the resonance half-width, then $A_m$ calculated using Equation \ref{MD10.22} is not a good estimation of the radial amplitude produced by the resonance because in this regime neighboring streamlines will cross and collisional dissipation cannot be ignored. 

In Figure \ref{resstrength} we display the forced amplitude on all 10 rings (inner and outer edges for the $\epsilon$ ring) due to all possible first-order Lindblad resonances of all Uranian moons out to Perdita. For the estimated mass of each moon, we use the radius measurements of \citet{2001Icar..151...69K} and \citet{2006Sci...311..973S} and consider a range of densities from $0.5$ to $1.3$ g/cm$^3$. In the left half of Figure \ref{resstrength} all resonances mentioned by \citet{1987AJ.....93..724P} are apparent in addition to a previously unexplored 2:1 ILR with Portia in the proximity of the $6$ ring. In the right side of Figure \ref{resstrength} we compare the amplitudes of the strongest resonances over a range of moon densities. The fainter patches in the left side of Figure \ref{resstrength} are due to resonances inducing much weaker amplitudes due to their large distance from the rings. Despite the separation in semi-major axis of the $\eta$ ring from the Cressida 3:2 ILR the $\eta$ ring is expected to be the most perturbed of all the Uranian rings in this framework. The next largest expected amplitudes are the Cordelia 24:25 OLR and the Ophelia 14:13 ILR that are thought to play a roll in shepherding the $\epsilon$ ring. If this is a realistic estimation of the strength of the resonances in the system, in the future we may be able to detect the $m=-24$ mode on the inner edge of the $\epsilon$ ring, which was previously detected by \citet{1995AAS...186.3302F} with occultation data and by \citet{2011epsc.conf.1224S} with images showing the ring's longitudinal brightness variations. Detecting the $\epsilon$ ring edge modes will first require determining the ring's edge positions and the removal of the larger amplitude $m=1$ normal mode which dominates its structure. Our analysis of these ring residuals as well as those for the other rings, whose structure is dominated by previously known normal modes, is ongoing and will be presented in a subsequent publication.


\section{Results} 

\begin{deluxetable}{lr}[t]
\tablecaption{$\eta$ ring $m=3$ best fit \label{parametertab}}
\tablewidth{0pt}
\tablehead{
\colhead{Parameter} &
\colhead{Final fit and scaled errors}
}
\startdata
$a$ (km) & $47176.447\pm0.086$ \\
$A_3$ (km) & $0.667\pm0.113$ \\
$\delta_3$ ($^{\circ}$) & $58.81\pm6.12$ \\
$\Omega_{p}$ ($^{\circ}/$day) & $776.58208\pm0.00169$ \\
\hline
$n_{Cressida}$ ($^{\circ}/$day) & $776.582789\pm0.000059$\tablenotemark{a} \\
$n_{Cressida}$ ($^{\circ}/$day) & $776.582414\pm0.000022$\tablenotemark{b} \\
$n_{Cressida}$ ($^{\circ}/$day) & $776.582447\pm0.000022$\tablenotemark{c} \\
\hline
$\chi^{2}$ & $13.861$ \\
$\chi^{2}/\nu$ & $0.308$ \\
$\sigma/\nu$ (km) & $0.555$ \\
$N$ & $49$ \\
$\#$ of parameters & $4$ \\
\enddata
\tablecomments{Listed on top are the four fit parameters and their formal 1-$\sigma$ errors resulting from our final fit, where we have assumed an error of $0.555$ km for each of the observed radii of the $\eta$ ring. We also list three published mean motions of Cressida for comparison with our pattern speed. The chi-squared and reduced chi-squared below are from the initial fit assuming an error of $1$ km for each radii. The unscaled errors of the parameters in the initial fit are roughly double the scaled errors from the final fit, in which we have used the standard deviation per degree of freedom as a universal error in the observed radii. The degrees of freedom $\nu=N-\#$ of fit parameters.}
\tablenotetext{a}{From \citet{2006Sci...311..973S}}
\tablenotetext{b}{From \citet{1998AJ....115.1195J}}
\tablenotetext{c}{From \citet{1998AJ....115.1190P}}
\end{deluxetable}

\begin{figure}[t]
\epsscale{1.1}
\plotone{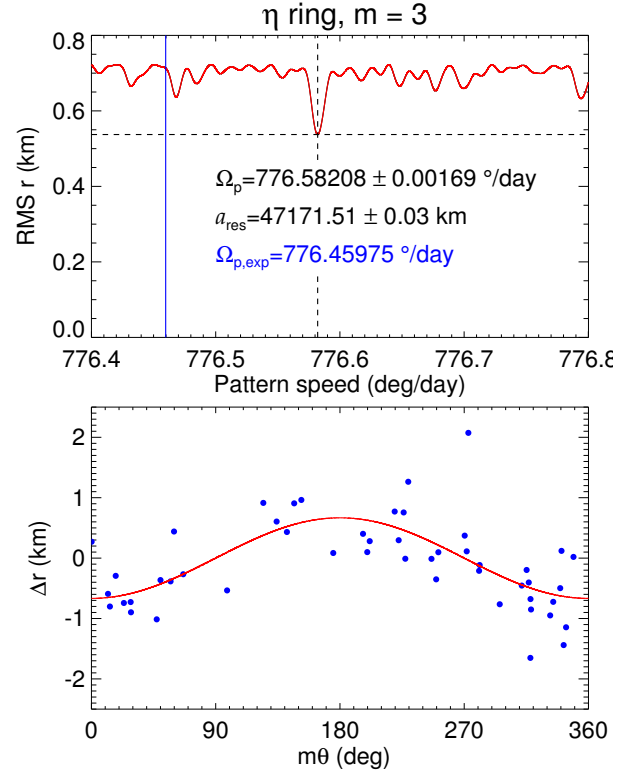}
\caption{The upper plot shows the RMS deviations of the observed radii, $r$, and the model fit (Equation \ref{modrad}) for a range of pattern speeds, $\Omega_{p}$. Listed in the top plot are the best fitting pattern speed and the corresponding radial location of the resonance, $a_{res}$. The blue line refers to the expected pattern speed for an $m=3$ normal mode oscillation. The lower plot shows the best fitting model (red line) and observed radii plotted vs. $m\theta=m[\lambda-\Omega_p(t-t_0)-\delta_m]$ after subtracting the semi-major axis of the ring ($\Delta r=r-a$). \label{etafit}}
\end{figure}

\begin{figure}[t]
\epsscale{1.1}
\plotone{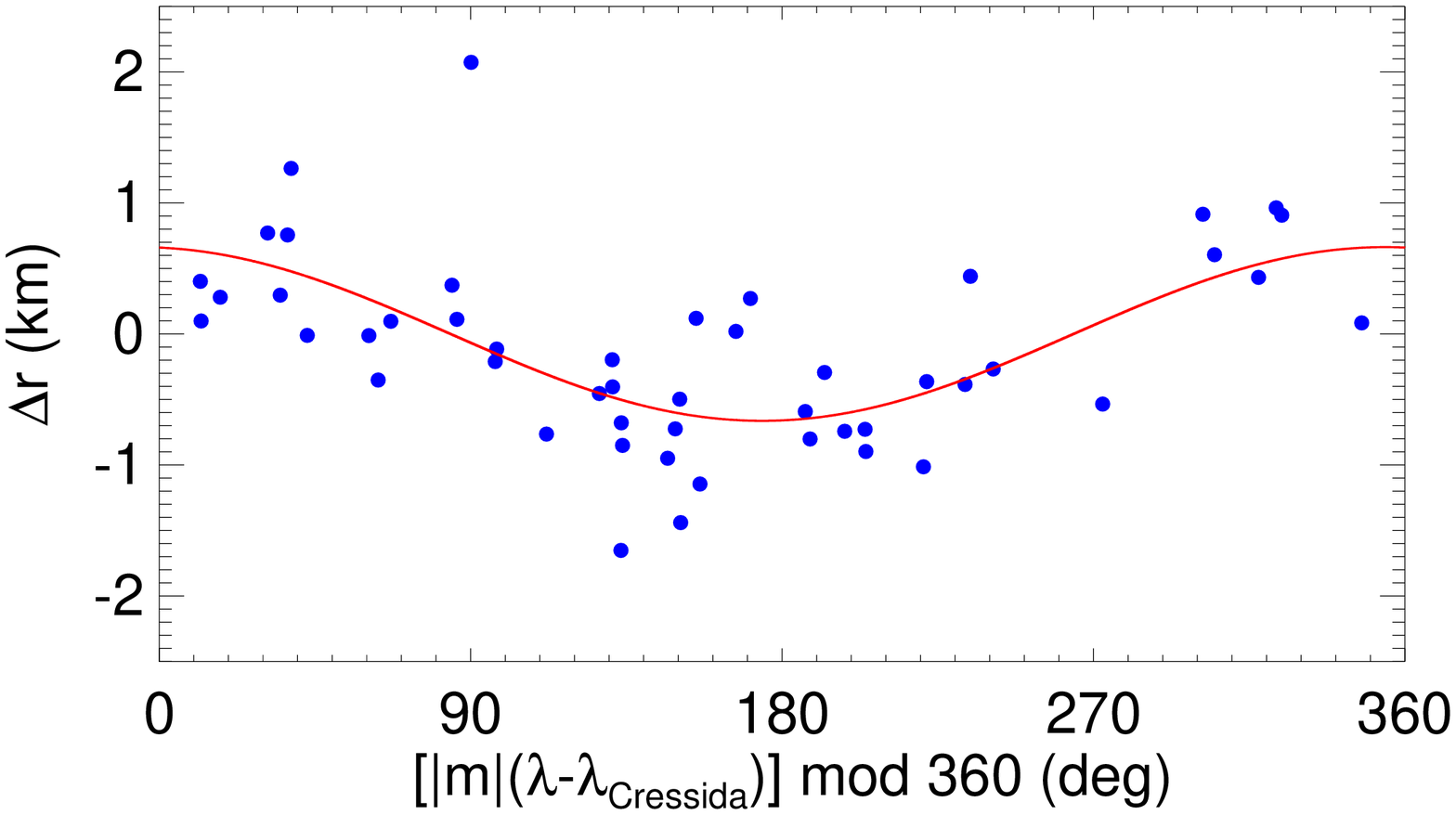}
\caption{This plot shows the $\eta$ ring structure in a reference frame tied to Cressida. One of the three outermost radial extents actually tracks Cressida, the others are located $\sim120^{\circ}$ apart. We obtained longitudes of Cressida at various times using the ura091.bsp and ura112.bsp SPICE kernels, available at \url{https://naif.jpl.nasa.gov/pub/naif/generic_kernels/spk/satellites/}. \label{cresalign}}
\end{figure}

After searching mode values from $m=-25$ to $25$ of all the rings, the strongest new feature we've found is an $m=3$ structure of the $\eta$ ring consistent with the expectations discussed above. In Figure \ref{etafit} we show the shallow minimum in RMS for our $\eta$ ring $m=3$ fits. The top plot shows the RMS deviations of the model radii from the observed radii at each pattern speed for $m=3$, zoomed in on the minimum. Listed are the best fitting pattern speed, the semi-major axis of the Cressida 3:2 ILR, and the expected pattern speed for an $m=3$ normal mode marked by the dashed line. Note that the best fitting pattern speed and the expected pattern speed for the semi-major axis of the $\eta$ ring are offset because this is not a normal mode oscillation, but is instead the effect of a resonance with a satellite whose perturbations force the pattern speed to match the satellite's mean motion. We further refine our best fit solution and formal errors by applying the best fit parameters ($a$, $A_m$, $\delta_m$, and $\Omega_p$) as a set of starting parameters for MPFIT, a non-linear least squares fitting IDL function \citep{2009ASPC..411..251M}. We've initially assumed an uncertainty of $1$ km in each of the $49$ observed radii of the $\eta$ ring, but found a reduced chi-squared of $0.308<<1$. We fit again to obtain the listed errors using the standard deviation per degree of freedom ($\sigma/\nu$) as a rescaled uncertainty in our observed radii which better represents the error of these data. The bottom plot shows the best fitting model radius curve on top of the observed radial separations from the fit semi-major axis of the ring, $\Delta r=r-a$. We've listed the final fit parameters and chi-squared analysis in Table \ref{parametertab}.

The best fitting pattern speed for this mode, $776.58208$ $\pm0.00169$ $^{\circ}/$day, is strikingly close to the published mean motion of Cressida, the fourth moon from Uranus. Most recently \citet{2006Sci...311..973S} listed Cressida's mean motion as $776.582789\pm0.000059^{\circ}/$day. All three of the measurements of Cressida's mean motion listed in Table \ref{parametertab} are well within the uncertainty of our detected pattern speed, supporting the proposed connection between this $m=3$ structure of the $\eta$ ring and Cressida. 

To solidify that the $m=3$ structure is real and is a result of perturbations from Cressida, we have inspected the alignment of the structure with Cressida. In this case, the $\eta$ ring ($a=47176.447$) is located exterior to the resonance ($a_{res}=47171.51$), and the dynamical model predicts that one of the three outer radial extents should track the motion of Cressida. That is, as the $m=3$ structure and Cressida both rotate around Uranus at $n_{Cressida}\simeq\Omega_{p}$ one of the apoapses is constantly aligned with Cressida. This can be confirmed by noting that the $m=3$ structure has a phase offset $\delta_3=58.81\pm6.12^{\circ}$ (this is the longitude of one of the 3 periapsis), which is roughly $60^{\circ}$ offset from Cressida's longitude ($359.50^{\circ}$) at the epoch of the fit. We show this alignment more precisely in Figure \ref{cresalign}, where we have determined the offset of each occultation scan longitude relative to Cressida's longitude at the observation time, $|m|(\lambda-\lambda_{Cressida})$. The apoapse of the phase-wrapped structure lags the longitude of Cressida by only $6\pm11^{\circ}$ (Cressida's longitude is $0^{\circ}$ and the fit sinusoid's largest radial excursion occurs at $354^{\circ}$). This suggests that the perturbations on the $\eta$ ring are due to its proximity to the 3:2 ILR with Cressida.

\begin{deluxetable*}{cccccc}[t]
\tablecaption{Mass and Density of Cressida \label{density}}
\tablewidth{0pt}
\tablehead{
\colhead{$A_3$ (km)} &
\colhead{Radius (km)} &
\colhead{$a$ (km)} &
\colhead{$a_{res}$ (km)} &
\colhead{$m_{Cressida}$ (kg)} &
\colhead{$\rho_{Cressida}$ (g cm$^{-3})$}
}
\startdata
$0.667\pm0.113$ & $41\pm2$ & $47176.447\pm0.086$ & $47171.51\pm0.03$ & $2.5\pm0.4\times10^{17}$ & $0.86\pm0.16$\\
\enddata
\tablecomments{We list the variables needed to solve for the mass of Cressida using Equation \ref{MD10.22}. For the calculation of $m_{Cressida}$ we used $GM_{Uranus}=5793951.3\pm4.4$ km$^3$ s$^{-2}$ from \citet{2014AJ....148...76J} and $G=6.67408\pm31\times10^{-11}$ m$^3$ kg$^{-1}$ s$^{-2}$ from \url{http://physics.nist.gov/cgi-bin/cuu/Value?bg}. Also note $\frac{2\alpha|f_d|}{j-1}\simeq1.545$ when $j=m=3$ for the case of the Cressida 3:2 ILR. The listed radius needed to calculate the density of Cressida comes from Voyager 2 photometry \citep{2001Icar..151...69K}.}
\end{deluxetable*}

Perhaps the most significant result of this work, shown in Table \ref{density}, is a determination of Cressida's mass using Equation \ref{MD10.22}. Given $A_3=0.667\pm0.113$ km we find $m_{Cressida}=2.5\pm0.4\times10^{17}$ kg. We use the effective radius for Cressida of $41\pm2$ km from \citet{2001Icar..151...69K} to calculate a density of $0.86\pm0.16$ g/cm$^3$ for Cressida. 

For our purposes, the $\eta$ ring is outside the width of Cressida's 3:2 ILR and the resulting estimation of $A_m$ is reasonable, but we note that this is not necessarily the case for all of the other rings and resonances. Curious readers should note, to test the applicability of Equation \ref{MD10.22}, we've calculated a resonance half-width of $\sim3.5$ km for Cressida's 3:2 ILR using Equation 10.23 from \citet{1999ssd..book.....M} along with our newly determined mass of Cressida. The other relevant variable inputs can be found in Tables \ref{parametertab} and \ref{density}. This half-width is less than the $5$ km separation of the resonance and ring, confirming we are justified in using Equation \ref{MD10.22}. Note that the $\sim1$ km resonance half-width quoted in the introduction was estimated by \citet{1987AJ.....93..724P} and results from an approximation of the resonance half-width equation as well as a different satellite mass.


\section{Discussion}
Since the Voyager 2 flyby of Uranus in 1986, several dynamicists have explored the stability of the inner Uranian moons. The moons Bianca, Cressida, Desdemona, Juliet, Portia, Rosalind, Cupid, Belinda, and Perdita are members of the most tightly packed system of moons in our solar system. Nicknamed the `Portia group' for their largest member, these satellites are thought to be unstable on short timescales compared to the age of the solar system. The stability of the Portia group is known to be highly sensitive to the masses of the individual satellites \citep{2015AJ....149..142F}, which are not well constrained. In fact, the mass we provide for Cressida is the first direct measurement of an inner Uranian satellite's mass. Past simulations \citep{1997Icar..125....1D,2012Icar..220..911F,2014MNRAS.445.3959Q,2015AJ....149..142F} have relied on treating a range of possible masses for the inner Uranian satellites and suggest that Cressida will cross orbits with Desdemona in under $10^6$ years \citep{2012Icar..220..911F}, given our mass density. Incorporation of our mass for Cressida should further constrain the timescale of satellite orbit crossing (collisions) and allow a future work to determine the masses of some of the other satellites through their resonant interactions. Strictly speaking, our density measurement does not necessarily represent a common density of the inner moons. However, a lower average satellite density will generally result in collisions occurring in the more distant future. 

\begin{figure}[t]
\epsscale{1.1}
\plotone{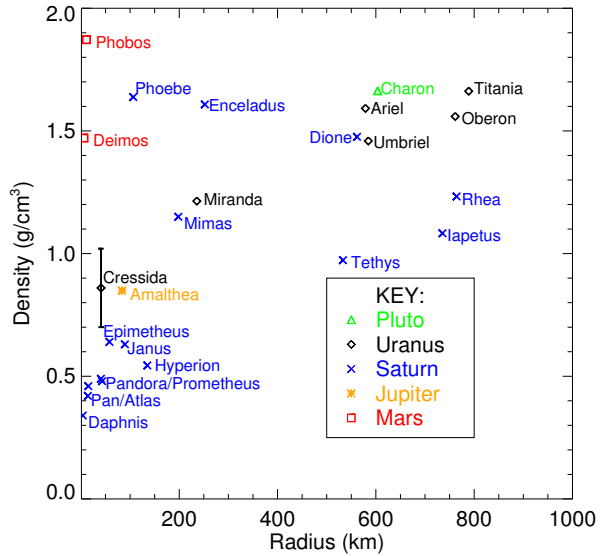}
\caption{The known densities of selected satellites in the solar system are plotted versus their radii. We represent the individual moons associated with particular planets using the point styles and colors labeled in the lower right of the plot. \label{satdensities}}
\end{figure}

\citet{2001Icar..151...51K} and \citet{2003AJ....126.1080D} detected a possible water ice absorption feature in Hubble Space Telescope near-infrared photometry of the largest inner moon Puck. Combining this with the previously mentioned size estimates has formed the presumption that Cressida and the other inner Uranian moons are likely composed of mostly water ice with at least a veneer or contamination of dark material to explain their low albedo and flat gray spectra. The range in densities of the larger Uranian moons, determined from mass \citep{1992AJ....103.2068J} and radius \citep{1988Icar...73..427T} measurements, have provided a presumed upper limit on the densities of the inner moons, usually with reference to the least dense major moon Miranda ($1.214\pm0.109$ g$/$cm$^3$)\footnote{\url{https://ssd.jpl.nasa.gov/?sat\_phys\_par}}. In Figure \ref{satdensities} we plot our average density of Cressida versus radius along with other satellites in our solar system, after \citet{2006Icar..185..258H}. Cressida is about $50\%$ denser than the inner icy moons of Saturn with comparable radii. It may be that Cressida, and the Uranian rings/moons in general, have either a lower porosity than these Saturnian analog or they have higher amounts of non-icy contaminants, as inferred by \citet{2013ApJ...765L..28T}. The contamination of denser and darker material may not be as high as previously expected, but it is substantial regardless.

This analysis shows that there is still information about Uranus' rings and moons  found in historical and ground based data. Still, the best means of obtaining the Uranian moon masses and compositions, determining the ultimate fate of the Portia group, and understanding the intricate structure of the rings is of course a Uranus orbiter mission.


\acknowledgments
We would like to thank Phil Nicholson for his insights regarding ring occultation observations and both Phil Nicholson and Pierre-Yves Longaretti for several fruitful discussions concerning the forced radial amplitudes of ring particles orbiting near Lindblad resonances. We'd also like to thank our anonymous reviewer for helpful suggestions and comments, ultimately improving the clarity of this work. This work was supported by the NASA Solar System Workings program grant NNX15AH45G.

\appendix
Included below are the occultation observation data we used in this analysis of the $\eta$ ring. The precise numbers for the ring's position are derived from an analysis of the entire Uranian ring data set, including re-determined pole position (Pole right ascention $=77.3105814^{\circ}$ and declination $=15.1697826^{\circ}$), standard gravitational parameter ($GM=5.793956433\times10^6$ km$^3$s$^{-2}$), gravitational harmonics ($J_2=3.340656\times10^{-3}$ and $J_4=-3.148536\times10^{-5}$), and time offsets. The numbers therefore can deviate slightly from previously published values. We list each observation ID, observing location, ring plane intercept time of the relevant electromagnetic wave observed, detected mid-radius of the $\eta$ ring, $m=3$ fit residuals, longitude of the observation, longitude of Cressida at this time, and reference to publications including the observation. Longitudes are measured in the prograde direction from the ascending node of Uranus' equator on the Earth's equator of the J2000 epoch. \citet{1988Icar...73..349F} have included all observations from 1977-1986 in their most recent fit, but more recent observations are unpublished. 
\startlongtable
\begin{longrotatetable}
\begin{deluxetable*}{cccccccc}
\tablecaption{$\eta$ ring occultation observation geometry \label{occdata}}
\tablewidth{0pt}
\tablehead{
\colhead{ID} &
\colhead{Observing Location} &
\colhead{Star Name Catalog ID} &
\colhead{$t_{\rm ring\ intercept}$ (UTC)} &
\colhead{$r$ (km)} &
\colhead{$r-r_{\rm fit}$ (km)} &
\colhead{$\lambda (^{\circ})$} &
\colhead{Reference}
}
\startdata
$1$ & Kuiper Airborne Obs. & U0 Hipparcos 71567 & 1977 MAR 10 17:48:26.95 & $47177.352$ & $ 0.346$ & $ 36.68$ & \citet{1977Natur.267..328E}\\
$2$ & Kuiper Airborne Observatory & U0 Hipparcos 71567 & 1977 MAR 10 19:20:03.28 & $47176.465$ & $ 0.673$ & $153.51$ & \citet{1977Natur.267..328E}\\
$3$ & Cerro Las Campanas Obs. & U5 UCAC2 25775788 & 1978 APR 10 03:00:16.19 & $47178.519$ & $ 2.107$ & $ 46.24$ & \citet{1978AJ.....83.1240N}\\
$4$ & Cerro Las Campanas Obs. & U5 UCAC2 25775788 & 1978 APR 10 03:50:16.64 & $47177.359$ & $ 0.535$ & $143.71$ & \citet{1978AJ.....83.1240N}\\
$5$ & Cerro Tololo Interamerican Obs. & U12 UCAC2 25096598 & 1980 AUG 15 22:20:37.70 & $47176.235$ & $-0.087$ & $ 22.05$ & \citet{1981AJ.....86..444E}\\
$6$ & European Southern Obs. 1-m & U12 UCAC2 25096598 & 1980 AUG 15 19:24:18.97 & $47175.681$ & $-0.476$ & $171.95$ & \citet{1981AJ.....86..444E}\\
$7$ & European Southern Obs. 1-m & U12 UCAC2 25096598 & 1980 AUG 15 22:20:36.26 & $47176.330$ & $ 0.013$ & $ 22.18$ & \citet{1981AJ.....86..444E}\\
$8$ & Anglo-Australian Telescope & U13 Hipparcos 77434 & 1981 APR 26 16:45:45.26 & $47176.726$ & $-0.341$ & $ 26.90$ & \citet{1982Natur.298..827F}\\
$9$ & Anglo-Australian Telescope & U13 Hipparcos 77434 & 1981 APR 26 17:53:41.63 & $47176.877$ & $-0.092$ & $163.55$ & \citet{1982Natur.298..827F}\\
$10$ & European Southern Obs. 2-m & U14 Hipparcos 79085 & 1982 APR 22 00:27:11.11 & $47175.767$ & $-0.184$ & $164.12$ & \citet{1986Icar...67..134F}\\
$11$ & Cerro Las Campanas Obs. & U14 Hipparcos 79085 & 1982 APR 21 23:07:45.20 & $47175.548$ & $-0.314$ & $ 24.83$ & \citet{1986Icar...67..134F}\\
$12$ & Cerro Las Campanas Obs. & U14 Hipparcos 79085 & 1982 APR 22 00:27:11.32 & $47174.793$ & $-1.158$ & $164.09$ & \citet{1986Icar...67..134F}\\
$13$ & Tenerife & U14 Hipparcos 79085 & 1982 APR 21 23:09:03.74 & $47176.060$ & $-0.027$ & $ 35.09$ & \citet{1986Icar...67..134F}\\
$14$ & Cerro Tololo Interamerican Obs. & U14 Hipparcos 79085 & 1982 APR 21 23:07:38.90 & $47175.718$ & $-0.143$ & $ 24.70$ & \citet{1986Icar...67..134F}\\
$15$ & Cerro Tololo Interamerican Obs. & U14 Hipparcos 79085 & 1982 APR 22 00:27:09.91 & $47175.595$ & $-0.353$ & $164.23$ & \citet{1986Icar...67..134F}\\
$16$ & Mt. Stromlo & U15 UCAC2 23648038 & 1982 MAY 01 13:53:02.53 & $47176.249$ & $ 0.275$ & $ 27.01$ & \citet{1986Icar...67..134F}\\
$17$ & Mt. Stromlo & U15 UCAC2 23648038 & 1982 MAY 01 15:01:45.77 & $47176.542$ & $-0.119$ & $162.76$ & \citet{1986Icar...67..134F}\\
$18$ & Mt. Palomar & U16 UCAC2 23892052 & 1982 JUN 04 02:50:34.77 & $47175.991$ & $-0.012$ & $ 32.37$ & \citet{1986Icar...67..134F}\\
$19$ & Mt. Palomar & U16 UCAC2 23892052 & 1982 JUN 04 03:50:47.86 & $47176.435$ & $-0.464$ & $156.73$ & \citet{1986Icar...67..134F}\\
$20$ & South African Astronomical Obs. & U17 Hipparcos 80841 & 1983 MAR 24 22:05:13.10 & $47176.847$ & $-0.238$ & $ 35.29$ & \citet{1986Icar...67..134F}\\
$21$ & Cerro Tololo Interamerican Obs. & U23 UCAC2 22735323 & 1985 MAY 04 02:28:59.92 & $47176.178$ & $-0.003$ & $318.94$ & \citet{1988Icar...73..349F}\\
$22$ & Cerro Tololo Interamerican Obs. & U23 UCAC2 22735323 & 1985 MAY 04 03:30:35.20 & $47176.885$ & $ 0.774$ & $229.93$ & \citet{1988Icar...73..349F}\\
$23$ & McDonald Obs. & U23 UCAC2 22735323 & 1985 MAY 04 03:37:42.72 & $47175.702$ & $-0.133$ & $221.66$ & \citet{1988Icar...73..349F}\\
$24$ & Cerro Tololo Interamerican Obs. & U25 UCAC2 22734194 & 1985 MAY 24 05:23:43.95 & $47176.565$ & $ 0.747$ & $316.13$ & \citet{1988Icar...73..349F}\\
$25$ & Cerro Tololo Interamerican Obs. & U25 UCAC2 22734194 & 1985 MAY 24 06:10:43.14 & $47176.151$ & $ 0.340$ & $233.85$ & \citet{1988Icar...73..349F}\\
$26$ & McDonald Obs. & U25 UCAC2 22734194 & 1985 MAY 24 05:22:06.85 & $47175.644$ & $-0.154$ & $326.24$ & \citet{1988Icar...73..349F}\\
$27$ & McDonald Obs. & U25 UCAC2 22734194 & 1985 MAY 24 06:17:47.01 & $47175.721$ & $-0.124$ & $223.28$ & \citet{1988Icar...73..349F}\\
$28$ & Mt. Palomar & U25 UCAC2 22734194 & 1985 MAY 24 05:23:00.51 & $47175.853$ & $ 0.059$ & $326.26$ & \citet{1988Icar...73..349F}\\
$29$ & Mt. Palomar & U25 UCAC2 22734194 & 1985 MAY 24 06:18:47.48 & $47175.497$ & $-0.360$ & $223.09$ & \citet{1988Icar...73..349F}\\
$30$ & IRTF & U28 UCAC2 22517254 & 1986 APR 26 10:58:06.57 & $47176.433$ & $-0.282$ & $333.24$ & \citet{1986Sci...233...79T}\\
$31$ & IRTF & U28 UCAC2 22517254 & 1986 APR 26 12:34:39.87 & $47175.910$ & $-0.632$ & $215.97$ & \citet{1986Sci...233...79T}\\
$32$ & IRTF & U1052 UCAC2 22296665 & 1988 MAY 12 10:56:55.69 & $47175.431$ & $-0.563$ & $293.09$ & \citet{1986Sci...233...65L}\\
$33$ & IRTF & U1052 UCAC2 22296665 & 1988 MAY 12 11:26:09.19 & $47176.094$ & $-0.584$ & $256.32$ & \citet{1988Icar...73..349F}\\
$34$ & IRTF & U83 UCAC2 22564036 & 1991 JUN 25 10:10:59.40 & $47176.545$ & $-0.530$ & $326.15$ & \citet{1988Icar...73..349F}\\
$35$ & IRTF & U83 UCAC2 22564036 & 1991 JUN 25 10:59:59.96 & $47176.530$ & $-0.581$ & $224.37$ & Unpublished\\
$36$ & IRTF & U84 UCAC2 22563790 & 1991 JUN 28 07:47:36.36 & $47176.741$ & $-0.197$ & $306.15$ & Unpublished\\
$37$ & IRTF & U84 UCAC2 22563790 & 1991 JUN 28 08:19:53.10 & $47175.300$ & $-0.507$ & $243.98$ & Unpublished\\
$38$ & Cerro Tololo Interamerican Obs. & U9539 UCAC2 23016546 & 1993 JUN 30 04:57:33.67 & $47176.082$ & $ 0.063$ & $351.62$ & Unpublished\\
$39$ & Cerro Tololo Interamerican Obs. & U9539 UCAC2 23016546 & 1993 JUN 30 05:54:55.13 & $47177.216$ & $ 0.256$ & $199.09$ & Unpublished\\
$40$ & South African Astronomical Obs. & U134 UCAC2 23509999 & 1995 SEP 09 15:29:49.47 & $47175.948$ & $ 0.126$ & $ 31.36$ & Unpublished\\
$41$ & South African Astronomical Obs. & U134 UCAC2 23509999 & 1995 SEP 09 16:57:52.04 & $47177.201$ & $ 0.292$ & $161.04$ & Unpublished\\
$42$ & IRTF & U137 UCAC3 141-413386 & 1996 MAR 16 11:59:43.68 & $47176.716$ & $ 0.936$ & $178.95$ & Unpublished\\
$43$ & IRTF & U137 UCAC3 141-413386 & 1996 MAR 16 12:43:23.55 & $47177.408$ & $ 0.372$ & $ 13.13$ & Unpublished\\
$44$ & Mt. Palomar & U138 UCAC2 24243463 & 1996 APR 10 09:27:44.07 & $47177.050$ & $ 0.139$ & $356.19$ & Unpublished\\
$45$ & Mt. Palomar & U0201 UCAC2 27214859 & 2002 JUL 29 07:20:09.62 & $47177.709$ & $ 0.829$ & $ 74.55$ & Unpublished\\
$46$ & Mt. Palomar & U0201 UCAC2 27214859 & 2002 JUL 29 07:29:17.80 & $47175.005$ & $-0.807$ & $117.00$ & Unpublished\\
$47$ & Voyager 2 - RSS &    & 1986 JAN 24 19:50:59.23 & $47176.817$ & $ 0.080$ & $343.08$ & \citet{1991uran.book..327F}\\
$48$ & Voyager 2 - RSS &    & 1986 JAN 24 22:44:28.11 & $47176.557$ & $ 0.375$ & $197.13$ & \citet{1991uran.book..327F}\\
$49$ & Voyager 2 - PPS & $\beta$ Per Hipparcos 14576 & 1986 JAN 24 19:36:54.98 & $47176.041$ & $ 0.132$ & $110.96$ & \citet{1991uran.book..327F}\\
\enddata
\tablecomments{The precise numbers for the ring's position are derived from an analysis of the entire Uranian ring data set, including re-determined pole position, $GM$, $J_2$, $J_4$, and time offsets. We used an epoch time, $t_0$, of UTC 1977 MAR 10 20:00:00.00 for all fits of this data set. The times, $t$, listed in this table refer to the exact time of the ray intercept in the ring plane for each occultation observation. The ring radii and longitudes are those observed at these times, where longitudes are measured in the prograde direction from the ascending node of Uranus' equator on the Earth's equator of the J2000 epoch. The residuals show separation of each observations' radii and the $m=3$ model radii.}
\end{deluxetable*}
\end{longrotatetable}

\acknowledgements

\bibliography{cressidaeta_revisions}
\bibliographystyle{aasjournal}

\end{document}